\documentclass[usenatbib]{mn2e}

\usepackage{amsmath,amssymb}
\usepackage{bm}
\usepackage{cases}
\usepackage{ulem}

\def\be{\begin{equation}}
\def\ee{\end{equation}}
\def\ba{\begin{eqnarray}}
\def\ea{\end{eqnarray}}
\def\nn{\nonumber}

\usepackage{color}              
\usepackage{graphicx}   
\usepackage{comment}
\usepackage{hyperref}
\hypersetup{
    colorlinks=true,
    linkcolor=red,
    citecolor=blue,
}

\newcommand\apj{Astro.~Phys.~J.}

\begin{document}

\title{
Search for CP Violating Signature of Intergalactic Magnetic Helicity in the Gamma Ray Sky
}

\date{\today}

\author[H. Tashiro, W. Chen, F. Ferrer, T. Vachaspati]{Hiroyuki Tashiro$^{\dag}$, Wenlei Chen$^{*}$, Francesc Ferrer$^{*}$,Tanmay Vachaspati$^{\dag}$\\
$^{\dag}$Physics Department, Arizona State University, Tempe, Arizona 85287, USA.\\
$^{*}$Physics Department and McDonnell Center for the Space Sciences, 
Washington University, St. Louis, MO 63130, USA.
}

\maketitle

\begin{abstract}
	The existence of a cosmological magnetic field could be revealed
	by the effects of non-trivial helicity on large scales.
We evaluate a CP odd statistic, $Q$, using gamma ray data obtained from 
Fermi satellite observations at high galactic latitudes to search for such
a signature. 
Observed values of $Q$ are found to be non-zero; the probability of a  similar signal in 
Monte Carlo simulations is $\sim 0.2\%$.
Contamination from the Milky Way does 
not seem to be responsible for the signal since it is present even for data
at very high galactic latitudes. Assuming that the signal is indeed due to a helical
cosmological magnetic field, our results indicate left-handed magnetic helicity
and field strength $\sim 10^{-14}~{\rm G}$ on $\sim 10~{\rm Mpc}$ scales. 
\end{abstract}

Parity (P) and charge conjugation (C) symmetry violating processes in the early
universe, such as during matter-genesis, may have produced a helical magnetic
field, with important implications for the structures we observe. In this case,
the observation of a cosmological magnetic field can probe the very early universe
($t \lesssim1~{\rm ns}$), provide information about particle physics at very high
temperatures ($T \gtrsim 1~{\rm TeV}$), and also characterize the cosmological
environment prior to structure formation.

Several tools to detect and study a cosmological magnetic field are known, including 
Faraday rotation of distant polarized sources and the cosmic microwave background (CMB), 
and the distribution of GeV gamma rays from TeV blazars (see \citet{Durrer:2013pga} for
a recent review). However, there are very few ideas for how to directly measure the helicity 
of a magnetic field \citep{Kahniashvili:2005yp,Tashiro:2013bxa}. The helicity of a magnetic 
field may be viewed as due to the screw-like (or linked) 
distribution of magnetic field lines. More formally, the magnetic helicity density within a
large volume $V$ is defined as
\begin{equation*}
 h = {1 \over V} \int_V  d^3 x ~ {\bm A} \cdot  {\bm B},
\end{equation*}
where $\bm A$ is the electromagnetic potential of magnetic field, 
${\bm B } = \nabla \times {\bm A}$. Magnetic helicity is odd under combined charge 
conjugation plus parity (CP) transformations. 

Indirect measures of magnetic helicity rely on measuring the non-helical power spectrum
and then deducing properties of the helical spectrum on the basis of MHD evolution~\citep{Christensson:2002xu,Campanelli:2004wm,Banerjee:2004df,Campanelli:2007tc,Boyarsky:2011uy,Kahniashvili:2012uj},
or else by constructing parity odd cross-correlators of CMB temperature and polarization~\citep{Caprini:2003vc,Kahniashvili:2005xe,Kunze:2011bp}.
Direct measures can only rely on the propagation of charged particles through
the magnetic field as these sample the full three dimensional distribution of the
field. Thus cosmic rays are sensitive to magnetic helicity \citep{Kahniashvili:2005yp}, 
as are GeV gamma rays that are produced due to cascades originating from TeV 
blazars \citep{Tashiro:2013bxa}. In the latter process, the original TeV
photon produces an electron-positron pair by scattering with extragalactic
background light (EBL) in a cosmological void region. The charged pair then propagate in the
intervening magnetic field, and finally up-scatter CMB photons to produce GeV gamma 
rays. In the context of a single TeV source, observed GeV gamma rays then carry information 
about the helicity of the intervening magnetic field. 
A key point of the present paper, also alluded to in~\citet{Tashiro:2013bxa},
is that the observed {\it diffuse} gamma ray sky may also hold information
about the cosmological helical magnetic field and CP violation in the early 
universe.

\begin{figure}
  \begin{center}
   \includegraphics[width=1.0\columnwidth]{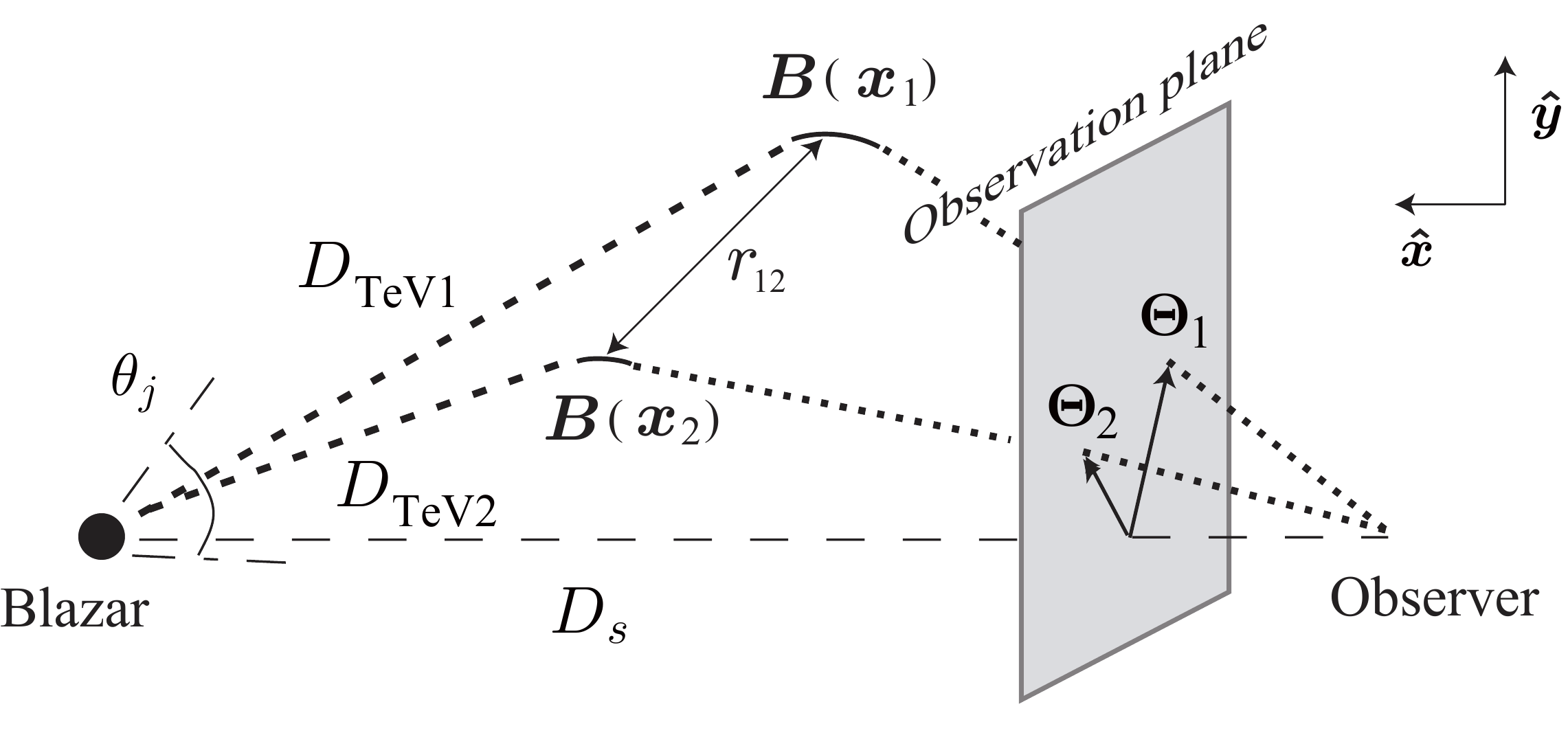}
  \end{center}
\caption{Events at two different energies sample the magnetic field
in regions of size $D_e \sim 30~{\rm kpc}$ (solid lines at the vertices
of the triangles). The regions themselves are separated by
distance $r$ which can be $\sim 100~{\rm Mpc}$ depending on the
energy difference of the two events. (Figure taken from~\citealt{Tashiro:2013bxa}.)}
\label{fig:cor}
\end{figure}

Assume we are located within the jet opening angle of a blazar but are off-axis (see 
Fig.~\ref{fig:cor}). 
A photon of energy $E_1\sim {\rm TeV}$ from the blazar propagates a distance
$D_{\rm TeV1} \sim 100~{\rm Mpc}$ and then scatters with an EBL photon to 
produce an electron-positron pair~\citep{Neronov:2009gh}.
The electron (positron) bends in the cosmological magnetic field and, after a typical
distance of about $30~{\rm kpc}$, up-scatters a CMB photon, that arrives
to the observer at the vectorial position ${\bm \Theta}_1$ in the observation plane. 
Similarly, another photon of energy $E_2$ arrives at ${\bm \Theta}_2$. Note that
the line-of-sight to the source defines the origin on the observation plane.

Let us define
$G(E_1, E_2) = \langle {\bm \Theta} (E_1)  \times {\bm \Theta} (E_2) \cdot \hat {\bm x} \rangle,$
where $\hat {\bm x}$ is perpendicular to the plane of observation and points towards
the source, and the ensemble
average is over all observed photons from the blazar. In~\citet{Tashiro:2013bxa} it
was shown that
\begin{align}
G(E_1, E_2) &\propto  {1 \over 2} M_H( |r_{12}| ) r_{12}.
\label{eq:g-2}
\end{align}
where $M_H$ is the helical correlation function of the
intervening magnetic field and defined by
\ba
\langle B_i ({\bm x} +{\bm r}) B_j ({\bm x})  \rangle = 
  M_N(r)  \left[ \delta_{ij} -{r_i r_j \over r^2} \right]+
M_L(r)  {r_i r_j \over r^2} &&\nn \\
&&
\hskip -2.5 cm
+ M_H (r) \epsilon_{ijl} r^l.
\ea
The distance $r_{12}$ in Eq.~(\ref{eq:g-2}) is given in terms of
the energies,
\begin{equation}
  r_{12} \approx D_{\rm TeV} (E_1) -D_{\rm TeV}(E_2)
  \label{rE1E2}
\end{equation}
with
\begin{equation}
 D_{\rm TeV} (E_{\rm TeV}) \sim 
 80 { \kappa \over (1+z_s)^2}  ~{\rm Mpc}~\left( {E_{\rm TeV} \over 10~{\rm TeV}}\right)^{-1},
 \label{eq:tev-meanfree}
\end{equation}
where $z_s$ is the redshift of the source and $\kappa$ is a parameter that depends on the 
EBL. We will take $1+z_s \sim 1$ and $\kappa \sim 1$~\citep{Neronov:2009gh}.
The overall proportionality factors in Eq.(\ref{eq:g-2}) depend on 
geometrical parameters such as the distance to the source and the energies, and will
not be important for what follows. Note that $r_{12}$ is positive if $E_1 < E_2$
because higher energy photons from the blazar produce electron-positron pairs more easily
and so $D_{\rm TeV}(E_1) > D_{\rm TeV}(E_2)$.

The correlation $G(E_1,E_2)$ is defined only if the TeV blazar is visible, since the
vectors ${\bm \Theta}$ originate at the location where the line of sight intersects the
observational plane. What if the TeV blazar is not visible? We can still measure
the helicity of an intervening magnetic field by noting that the highest energy
photons deviate the least from the source position. We can thus approximate 
the position of the blazar by the position of a photon with the highest energy
$E_3$ and the relevant correlator is 
\begin{equation*}
 G(E_1, E_2; E_3) = \langle ({\bm \Theta} (E_1) - {\bm \Theta} (E_3) ) \times 
                               ( {\bm \Theta} (E_2) - {\bm \Theta}(E_3) )
                                   \cdot {\hat {\bm x}_3} \rangle
\end{equation*}
and we will always assume the ordering $E_1 < E_2 < E_3$.
The vector ${\hat {\bm x}_3}$ points in the direction of the $E_3$ photon.

Diffuse gamma rays are observed on a sphere (the sky) and not on a plane and so
the statistic $G(E_1,E_2; E_3)$ needs to be modified suitably. We propose the
statistic (which is almost our final expression),
\begin{align*}
	Q'(E_1,E_2,E_3) & =  
\langle ({\bm n} (E_1)-{\bm n}(E_3)) \times \\
& \hskip 2 cm ({\bm n} (E_2)-{\bm n}(E_3)) \cdot {\bm n}(E_3)\rangle  \\
& = \langle {\bm n} (E_1) \times  {\bm n} (E_2) \cdot {\bm n}(E_3)\rangle,
\end{align*}
where ${\bm n}(E)$ denotes the (unit) vector to the location of the photon of energy
$E$ on the sky.

The problem with $Q'$ is that we cannot be sure that the photon of energy $E_3$
corresponds reasonably to a source for cascade photons. Also, in the case when
the TeV source was known, the ensemble average is taken over all cascade
photons {\it originating from the source}. In our case, we don't even know if there
is a TeV source, let alone which photons originate from a cascade and which do not.
However, if we work on the hypothesis that some of the photons that are not too
far away from the location of an $E_3$ photon originate from the same source and are possibly
due to a cascade, the statistic should still make sense if we restrict the average to
a region close to the location of the $E_3$ photon. (Note that such a 
region may contain photons unrelated to the $E_3$ cascade, but their 
contribution to the odd-statistic $Q$ will add up to zero on average.) 
To do this we can introduce a
window function that will preferably sample $E_1$ and $E_2$ photons close to
the chosen $E_3$ photon. The simplest implementation, and the one we have chosen,
is to use a top-hat window function with a radius that we treat as a free parameter. 
Further, we ensemble average over all $E_3$ photons since we do not
know if any given $E_3$ photon is due to a TeV source. Then, our final expression
for the statistic is 
\begin{align*}
Q &\left(E_1,E_2,E_3,R\right)  = \frac{1}{N_1N_2N_3} \times \\
	  &\sum_{i=1}^{N_1}\sum_{j=1}^{N_2}\sum_{k=1}^{N_3} 
                    W_R({\bm n}_i (E_1) \cdot {\bm n}_k (E_3))\, 
		    W_R({\bm n}_j (E_2) \cdot {\bm n}_k (E_3))\,\\
		    &\quad \quad \quad \quad \quad{\bm n}_i (E_1) \times  {\bm n}_j (E_2) \cdot {\bm n}_k (E_3),
\label{Qdef}
\end{align*}
where the indices $i,j,k$ refer to different photons and the top-hat window function
$W_R$ is given by
\begin{numcases}
{W_R(\cos\alpha) =}
1, \nonumber & \qquad for $\alpha \leq R$
\\
0,  & \qquad otherwise.
\end{numcases}

With a top-hat window function, the statistic can also be written as
\begin{equation}
Q(E_1,E_2,E_3,R)  = \frac{1}{N_3} \sum_{k=1}^{N_3} 
                                   {\bm \eta}_1\times {\bm \eta}_2 \cdot {\bm n}_k(E_3)
\label{Qdef2}
\end{equation}
where
${\bm \eta}_a = (1/N_a) \sum_{i \in D(n_k,R)} {\bm n}_i (E_a)$, $a=1,2$
and $D(n_k(E_3),R))$ is the ``patch'' in the sky with center at the location of 
${\bm n}_k(E_3)$ and radius $R$ degrees. Essentially, ${\bm \eta}_a$ are the
average locations of photons of energy $E_a$ within a patch, and $Q$ is given 
by the radial component of ${\bm \eta}_1 \times {\bm \eta}_2$ averaged over all
patches in the sky that are centered on photons with energy $E_3$. 

\begin{figure}
\includegraphics[width=\columnwidth]{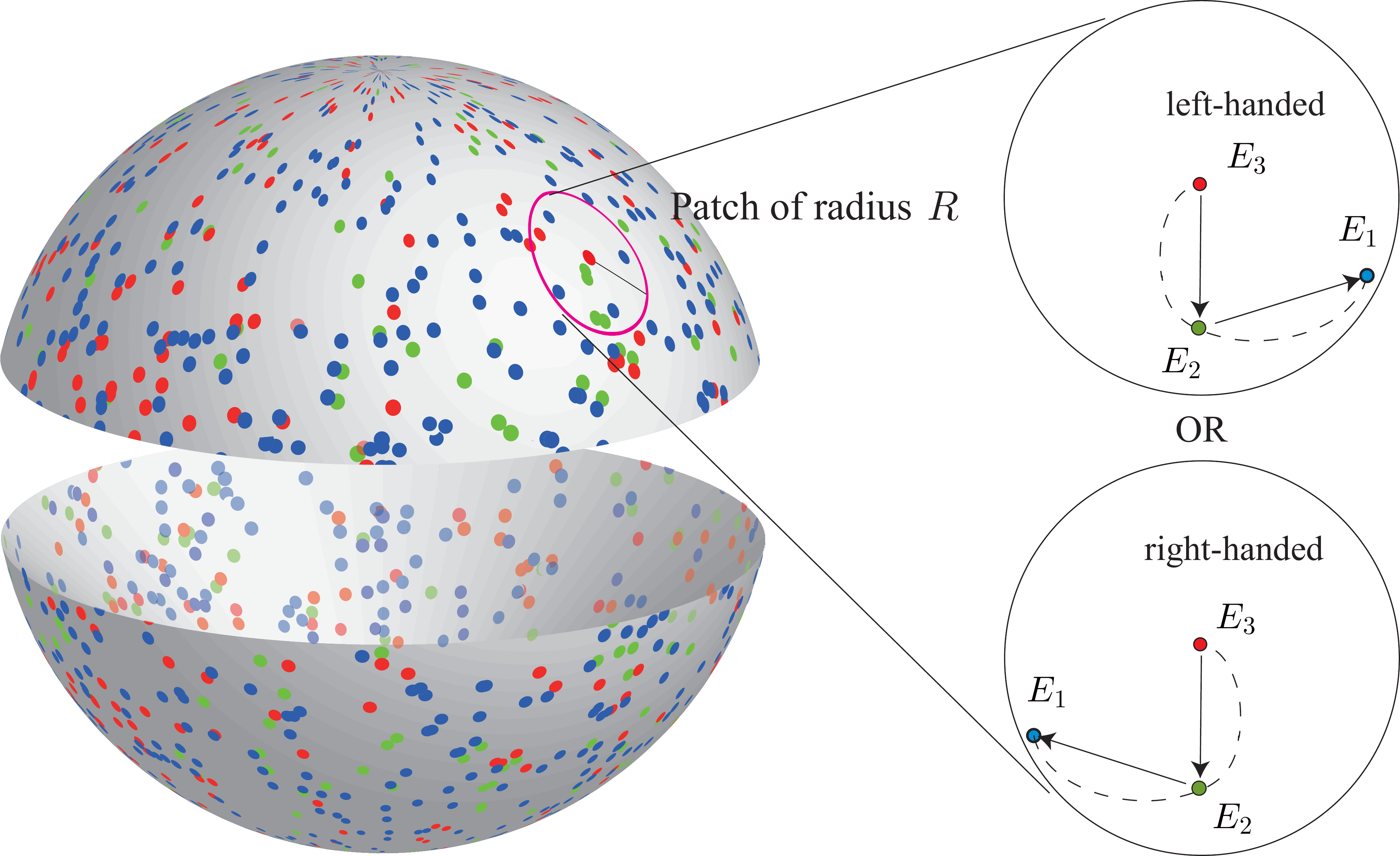}
\caption{
Illustration of the cut-sky with gamma rays distributed on it. Patches of radius $R$
degrees are centered on the highest energy gamma rays. In those patches we
test if the lower energy photons are distributed along left- or right-handed spirals.}
\label{fig:sphere}
\end{figure}

An intuitive picture for the meaning of the correlator is shown in Fig.~\ref{fig:sphere}.
We observe photons of three different energies (illustrated by three different colors)
on the cut-sky away from the 
galactic plane. We assume that the highest energy $E_3$ photons approximately
represent the source directions. Lower energy ($E_1$ and $E_2$) photons in patches of
some radius $R$ around the position of the $E_3$ photon are more likely to be from the 
same source. Then we consider the vectors in the patches as shown in
Fig.~\ref{fig:sphere} and ask if the directed curves from $E_3$ to $E_2$ to $E_1$ are
bent to the left or to the right, {\it i.e.} are the photons of decreasing energy
in patterns of left-handed or right-handed spirals? A positive (negative) value of the
statistic $Q$ implies that there is an excess of right-handed (left-handed) spirals in the
gamma ray sky. 

Next we measure the value of $Q$ on the 
emission detected by the
Fermi-LAT, using $\sim 60$ months of data.\footnote{Our analysis tools are available on the
wiki~\url{https://sites.physics.wustl.edu/magneticfields/wiki/index.php/Search_for_CP_violation_in_the_gamma-ray_sky}.} The data were processed with
the FERMI SCIENCE TOOLS\footnote{The Fermi Science Support Center~(FSSC),~\url{http://fermi.gsfc.nasa.gov/ssc}} to mask regions of the sky heavily
contaminated by Galactic diffuse emission and known point sources. We selected
LAT data from early-August 2008 through end of January 2014 (weeks 9 to 307) that
were observed at galactic latitudes, $|GLAT| \geq 50^\circ$. To ensure that the events are photons with high
probability, we use the Pass 7 Reprocessed data in the CLEAN event class.
Contamination from photons produced by cosmic-ray interactions in the upper
atmosphere is avoided by excluding events with zenith angles greater than
$100^\circ$, and only data for time periods when the spacecraft's
rocking angle was below $52^\circ$ were considered. Since
we are interested in the diffuse emission, we mask out a $3^\circ$ angular
diameter around each source in the First LAT High-Energy 
Catalog~\citep{TheFermi-LAT:2013xza}.

\begin{figure}
\centering
\begin{minipage}{.49\columnwidth}
	\centering
	\includegraphics[width=\linewidth]{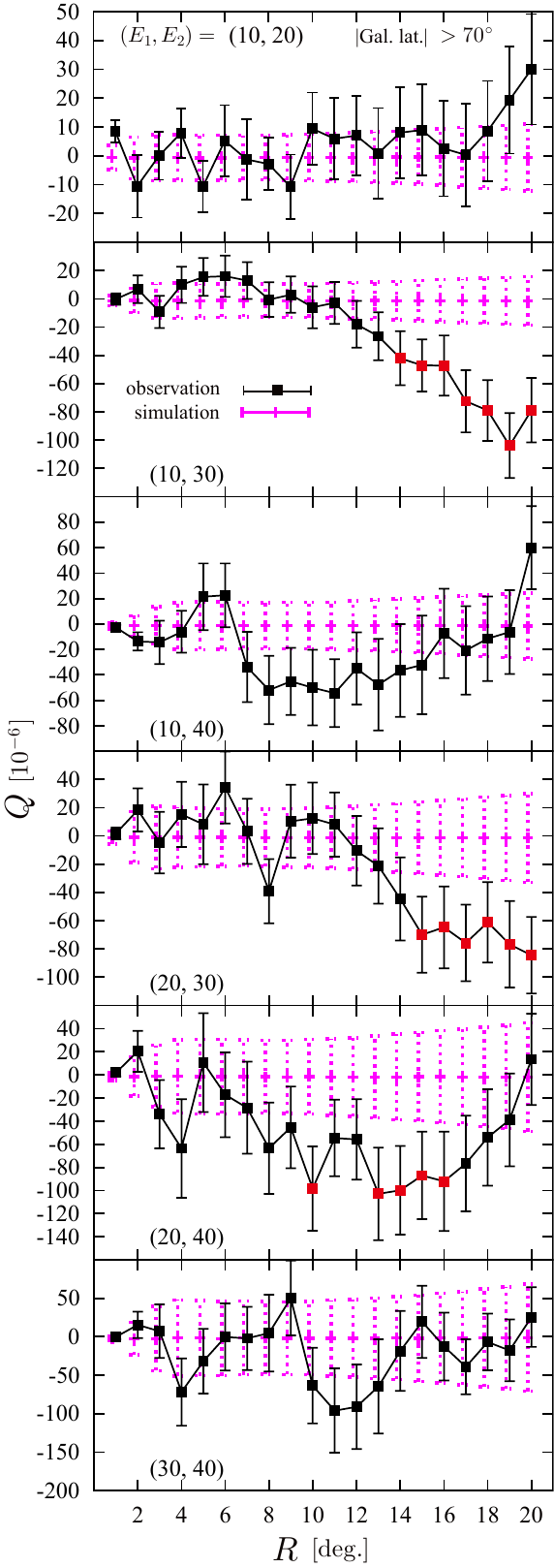}
\end{minipage}
\begin{minipage}{.49\columnwidth}
	\includegraphics[width=\linewidth]{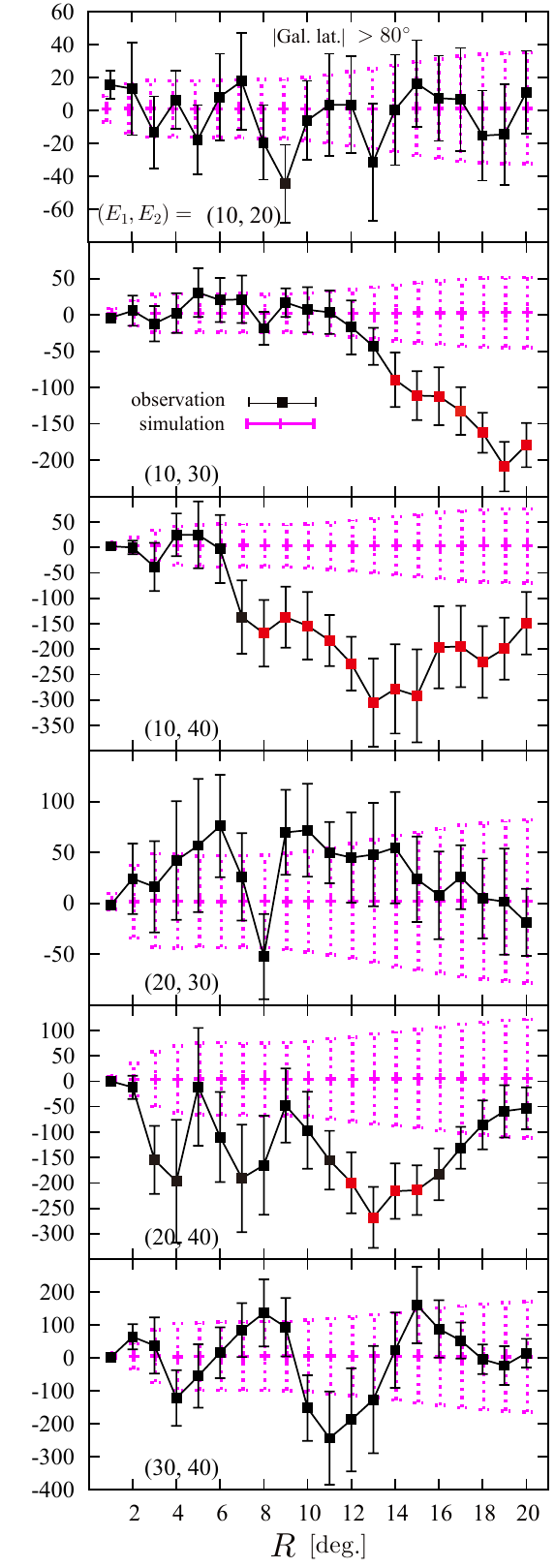}
\end{minipage}
	\caption{
$Q$ vs. patch radius in degrees for different combinations of
$\left\{ E_1,E_2 \right\} \in \left\{10,20,30,40\right\}~{\rm GeV}$
when patches centered on $E_3=50~{\rm GeV}$ photons are considered at absolute 
galactic latitude greater than $70^\circ$ (left column) and $80^\circ$ (right column). 
Also, shown are $1\sigma$ spreads (magenta error bars) obtained from simulated data. 
$Q$ values that are non-zero at greater than $2\sigma$ are shown by red squares
in the plots.
}
\label{QvsRclean}
\end{figure}

We restrict our analysis to the energy range $10-60$ GeV 
and we bin the
data in $5$ linearly spaced energy bins of width $\Delta E = 10$ GeV.
We will label events with energies in $(E, E+\Delta E)$ by $E$, 
e.g. $10$ GeV photons refers to data in the $(10,20)~{\rm GeV}$ bin. 
The total number of photons above $60^\circ$ absolute galactic latitude in each 
of the five bins of increasing energy is 7053, 1625, 726, 338 and 200.
We then evaluated $Q$ using Eq.~(\ref{Qdef2}) for patches of radius 
$R=1^\circ-20^\circ$ and for each of the six possible combinations of 
$E_1 < E_2 < E_3=50~{\rm GeV}$ as shown in~Fig.~\ref{QvsRclean}. 
The left and right columns display the analysis with $E_3=50~{\rm GeV}$ 
photons that are restricted to lie with absolute galactic latitude larger than $70^\circ$
and $80^\circ$ respectively.
For the smallest values of $R$, some of the patches centered on
the highest energy $E_3$ events will not contain any low-energy photon, 
and we set $Q=0$ in this case.
To each data point we associate the ``standard error'' bar, 
which is given by the standard deviation of the distribution of $Q$ 
values over different patches, $\sigma_Q$, divided by 
$\sqrt{N_3}$ where $N_3$ is the number of $E_3$ photons, 
which is the same as the number of patches. 
Thus, $\delta Q = \sigma_Q/\sqrt{N_3}$. 
We also evaluated errors due to the Fermi-LAT PSF\footnote{\url{http://fermi.gsfc.nasa.gov/ssc/data/analysis/documentation/Cicerone/Cicerone_LAT_IRFs/IRF_PSF.html}}.
We added (Gaussian) noise to 
the data consistent with the PSF in every energy bin. As the width of the PSF in the
lowest energy bin is $\sim 8~{\rm arcmin}$, these resolution errors are small, of
order 10\% of the standard error, and are not shown.
For comparison, we have generated synthetic data using a uniform distribution of 
gamma rays at each energy. 
Since we are only looking at the diffuse gamma ray background and have cut out
identified sources, a uniform distribution is a reasonable model.
The mean value of $Q$ and its standard deviation are evaluated over
$10^4$ realizations of synthetic data that are treated exactly like
the real data. 
As shown in Fig.~\ref{QvsRclean}, the mean value for the synthetic
data is zero as no CP violation is present. The $1\sigma$ spread obtained from 
the synthetic data, and the standard error obtained from real data are comparable.
To quote error bars we always take the larger of the two spreads.

Non-zero values of $Q$ at greater than $2 \sigma$ level occur for several
energy combinations and for different patch sizes. Most significantly,
the (10,40) energy combination plot in the right column shows $> 2\sigma$
deviations from zero for all patch sizes from $R=8^\circ-20^\circ$.
We should keep in mind, however, that we have scanned over
several parameters and the actual significance of our results should be 
modulated by a penalty factor discussed further below. 

\begin{figure}
\includegraphics[width=1.0\columnwidth]{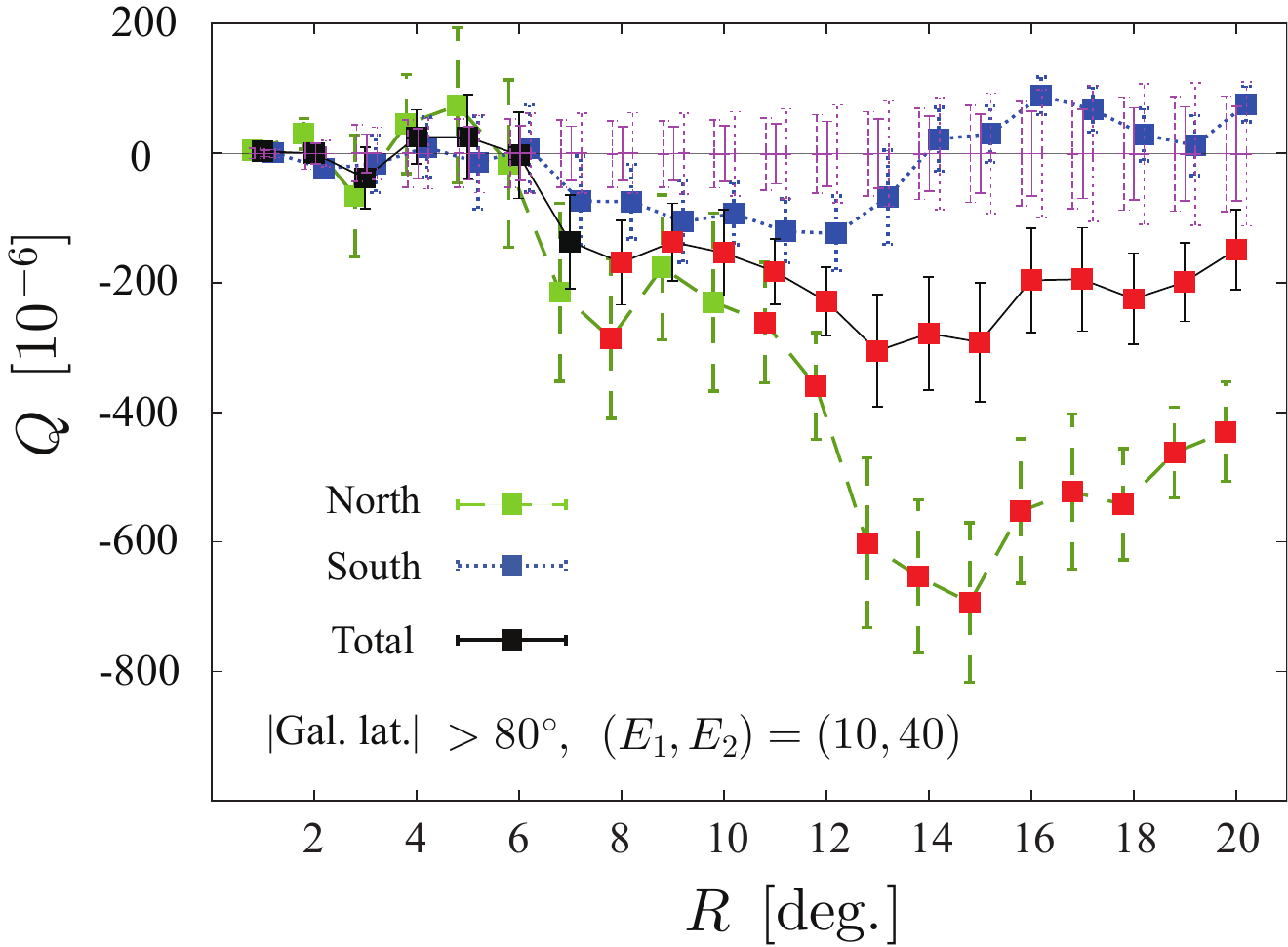}
\caption{
$Q$ vs. $R$ for the northern and southern hemispheres and for both combined.
The southern hemisphere $Q$ values are consistent with zero at the $\sim 2\sigma$
level; the northern
hemisphere values are non-zero even at the $\sim 3\sigma$ level for
larger patches.
}
\label{Qrns}
\end{figure}

When we analyze the (10,40) data separately for the northern and southern 
hemispheres, as in Fig.~\ref{Qrns}, we find non-zero $Q$ values with $\gtrsim 3 \sigma$
significance in the northern hemisphere with $R=11^\circ-20^\circ$. 
The signal in the southern hemisphere is marginally below the $2\sigma$ level.
One possible reason is that photon statistics in the south is 10-33\% poorer in the
four energy bins in the 10-50 GeV range. Another possible reason is that there genuinely is a north-south asymmetry
in the cascade photons, especially in the small polar regions we are considering.
As more data is collected, a clearer picture will emerge.

Our results have an interpretation in terms of cascade gamma rays originating from
TeV blazars in the presence of a cosmological magnetic field with helicity. 
However there are other possibilities too. We now discuss that the signal may be
due to contamination, or a statistical fluctuation, or perhaps a systematic error.

We have tested the possibility of Milky Way contamination by only considering 
patches centered at very high absolute galactic latitudes.
We find that the signal actually grows stronger if we restrict the patch centers to 
be at higher absolute galactic latitudes ($|b| > 80^\circ$ compared to $|b| > 70^\circ$). 
The stronger signal at high latitudes suggests that the effect is extragalactic.
In addition, if Milky Way contamination was responsible for the signal, the signal should
continue to grow for large $R$ since such patches extend to lower galactic latitudes. However, 
we see a peak structure at $R\sim 12^\circ$.

As alluded to above, scanning over several parameters might artificially 
bias the significance of the signal. To account for this so called 
``look elsewhere effect'' we estimate the ``penalty factor'' 
introduced by our scanning over angle and energy. We perform a Monte Carlo 
simulation, in which synthetic data is subject to the same analysis as
the real dataset, and count the occurrence of $Q$ values that deviate by more 
than $2\sigma$ for 13 consecutive values of patch radii, $R$, 
in {\it any} energy bin, $(E_1,E_2)$, and with cuts of 
$|b| > 70^\circ,\,80^\circ$. Such signals only appear with probability 
$\sim 0.002$.
As more data is accumulated, our findings will become more robust as 
we will have smaller error bars. In addition, since the signal should also 
appear in future data, we will be able to confirm our positive findings at 
patch radius $\Theta \approx 12^\circ$ without the need for scanning over 
parameters.

Systematic errors may be present in the data sets we have used for
some unknown experimental reasons. These are difficult to track down but it
makes sense to ask what systematic transformations of the data might eliminate
the signal. Since $Q$ is a (pseudo) scalar, it is unaffected by an overall rotation.
If we could rotate only photons in one energy bin and in each individual patch
around the axis through the center of the patch, we may be able to undo the 
signal. However, such a rotation on the data is not possible because there are 
many overlapping patches on the sky and the rotation cannot be defined for 
photons in the regions common to two distinct patches. A systematic transformation we
have investigated is a rotation of the $10-20~{\rm GeV}$ photons about the
polar axis, and in opposite senses in the northern and southern hemispheres. 
The transformation shifts the azimuthal angles of only the $10~{\rm GeV}$ bin 
by an angle $\alpha$ in the northern hemisphere, and by $-\alpha$ in the southern
hemisphere, where $\alpha$ is varied in steps of 10 arcminutes in the interval
$(-0.5^\circ,+0.5^\circ)$. However, we find that the value of $Q$
remains unchanged by these rotations. If the signal is due to some
other systematic, these need to be quite complicated as the photons at different 
energies need to be shifted with respect to each other in a parity odd way, and
in such a way that the signal does not re-appear in the energy combinations where
it is currently absent. 
In a preliminary analysis, we have used the Fermi time exposure
data to perform Monte Carlo simulations and still find the signal to be
significant. We are currently exploring other tests.

Next we assume that the signal is indeed due to the cascade process in the
presence of a cosmological magnetic field. What properties of the magnetic
field can we deduce from the results?

We can estimate the magnetic field strength if we assume that the patch radius
at which we get a signal is determined by the bending of cascade electrons in the 
magnetic field. The bending angle is estimated as \citep{Tashiro:2013bxa}
\begin{equation*}
 \Theta (E_\gamma) \approx
 7.3 \times 10^{-5} ~ \left( { B  \over 10^{-16} {\rm G}}\right) 
 \left( {1 {\rm Gpc}} \over D_s \right) \left( {E_\gamma
  \over 100 {\rm GeV}}\right)^{-3/2} .
\end{equation*}
With $\Theta \approx 12^\circ$, $E_\gamma \approx 10~{\rm GeV}$,
$D_s \approx 1000~{\rm Mpc}$, we obtain $B \sim 10^{-14}~{\rm G}$. 
This value is about two orders of magnitude larger than the lower bound found 
in~\citet{Neronov:1900zz} and consistent with the claimed
measurement in~\citet{Ando:2010rb} and~\citet{Essey:2010nd} (also see \citealt{Neronov:2010bi}).
In this connection we should point out that there is debate on whether pair produced 
electrons and positrons isotropize due to plasma instabilities~\citep{Tavecchio:2010mk,Dolag:2010ni,Broderick:2011av,Miniati:2012ge,2012ApJ...758..102S}
or if their propagation is simply given by bending due to a Lorentz force. 
Our results favor the latter scenario as it is hard to see how a plasma 
instability could give rise to a CP violating signature of the type we find. 

The energy combinations $(E_1,E_2)$ determine the distance on
which the gamma rays probe the magnetic field correlation function. 
From Eq.~(\ref{rE1E2}) with $z_s\sim 1$ and the relation for the observed
gamma ray energy,
$(E_{\gamma}/88~{\rm GeV})^{1/2} \sim E_{\rm TeV}/10~{\rm TeV}$ ~\citep{Neronov:2009gh},
we find that the (10,40)~GeV combination of energies probes distances 
$\sim 10~{\rm Mpc}$.
This should be considered as an order of magnitude estimate since we 
cannot be sure of the parameters $\kappa$ and $z_s$ that enter Eq.~(\ref{rE1E2}),
and also the relation (\ref{rE1E2}) was only derived in the case that the $E_3$ photon
points back to the source.

The results in Fig.~\ref{QvsRclean} show a strong CP violating signal in the 
$(10,40)~{\rm GeV}$ panel, less strong signals in the $(10,30)$ and 
$(20,40)$~GeV cases, but not in other energy panels.
One possible reason is that we did not detect cascade
photons from the same source in all energy bins.
Our CP violating signal arises in the energy combination $(E_1, E_2, E_3)$ when
cascade photons with energy $E_1$, $E_2$ and $E_3$ come from the same
source in the same patch. Since the sources of diffuse
gamma rays are unresolved, this suggests that the TeV blazars that
source cascade photons are very far and therefore 
the fluxes of cascade photons are very low. There is a possibility that
we have observed cascade photons from the same source in $E_1$, $E_2$
and $E_3$ energy bins but have not yet detected photons in the $E_1'$ bin. 
If this is the case, the CP violating signal will be present in the energy combination 
$(E_1, E_2, E_3)$ but not in $(E_1', E_2, E_3)$. Besides, TeV blazars also
emit GeV photons directly. Since the photon flux of blazars has a red spectrum, 
these direct GeV photons from unresolved blazars can dominate cascade photons 
in diffuse gamma rays. This contamination due to direct GeV photons can reduce 
the CP violating signals. 

The appearance of the CP violation signal only in the $(10,40)$~GeV
panel can also be explained in terms of magnetic field structures.
The connection between 
magnetic field helicity and correlators of gamma ray arrival directions given 
in Eq.~(\ref{eq:g-2}) only holds for identified blazars. A more detailed analysis
for the diffuse gamma ray flux, though with several simplifying assumptions,
shows the correspondence
\ba
Q &=& a(E_1,E_2,R) M_H (r_{12}) + a(E_2,E_3,R) M_H(r_{23}) \nn \\ 
&& +  ~a(E_3,E_1,R) M_H(r_{31}) 
\ea
where $a(E,E',R)$ is a function of the photon energies and the patch size.
Thus the signal seen in a panel depends on the details of the magnetic helicity 
spectrum at several different length scales, and on combinations of the other 
parameters. In principle,
the signal in the various panels can help us reconstruct the magnetic helicity spectrum,
though this will require more detailed investigations, some of which are under
way \citep{TashiroVachaspati2014}. 

Finally, since we find $Q < 0$, this indicates that the cosmological magnetic field has 
left-handed helicity. This could be very interesting for particle physics and early universe 
cosmology since baryogenesis, which requires fundamental CP violation, predicts 
magnetic fields with left-handed helicity~\citep{Vachaspati:2001nb},
while leptogenesis predicts right-handed helicity~\citep{Long:2013tha}. 
Inflationary models that produce helical magnetic fields have also been proposed
\citep{Caprini:2014mja} and can be distinguished from matter-genesis models by the 
spectral features of the magnetic fields they produce.

\medskip
We thank Roger Blandford, Jim Buckley, Nat Butler, Lawrence Krauss, Alexander Kusenko, 
Owen Littlejohns, Andrew Long and Stefano Profumo for helpful comments. TV is grateful to
the IAS, Princeton for hospitality where some of this work was done.
We are also grateful for computing resources at the ASU Advanced Computing Center (A2C2).
This work was supported by the DOE at ASU and at WU.


\end{document}